\documentclass[12pt]{article}
\usepackage{amsmath}
\usepackage[superscript,biblabel]{cite}
\usepackage{setspace}
\usepackage{authblk}
\usepackage{graphicx}
\usepackage{rotating}
\usepackage{enumerate}
\usepackage{xurl}
\usepackage{pdflscape}
\usepackage{booktabs}
\usepackage{layout}
\usepackage{caption}
\usepackage{subcaption}

\usepackage[utf8]{inputenc}

\usepackage[ruled,vlined]{algorithm2e}
\usepackage{bm}
\usepackage{psfrag,epsf}
\usepackage{float}
\usepackage{ amssymb }
\usepackage[table,xcdraw]{xcolor}
\usepackage{enumitem}
\usepackage{hyperref}
\usepackage{url}
\usepackage{scalerel}
\usepackage{lscape}
\usepackage{booktabs}

\def\spacingset#1{\renewcommand{\baselinestretch}%
{#1}\small\normalsize} \spacingset{1}

\title{\bf 
Modeling Alzheimer’s Disease: Bayesian Copula Graphical Model from Demographic, Cognitive, and Neuroimaging Data
}
\author[1]{Lucas Vogels}
\author[1]{Reza Mohammadi}
\author[1]{Marit Schoonhoven}
\author[1]{Ş. İlker Birbil}
\author[2]{Martin Dyrba}
\author[$\text{}$]{for the ADNI\footnote{Data used in preparation of this article were obtained from the Alzheimer’s Disease Neuroimaging Initiative (ADNI) database (\url{https://adni.loni.usc.edu}). As such, the investigators within the ADNI provided data but did not participate in analysis or writing of this report. A complete listing of ADNI investigators can be found at: \url{https://adni.loni.usc.edu/wp-content/uploads/how_to_apply/ADNI_Acknowledgement_List.pdf}}}
\affil[1]{Amsterdam Business School, University of Amsterdam, Amsterdam, the Netherlands}
\affil[2]{German Center for Neurodegenerative Diseases (DZNE), Rostock, Germany}

\date{\parbox{\linewidth}{\centering
  \today\endgraf\bigskip}
  \parbox{\linewidth}{\scriptsize{\textbf{Corresponding author:} Lucas Vogels, Amsterdam Business School, Plantage Muidergracht 12, 1018 TV Amsterdam, the Netherlands, email: l.f.o.vogels@uva.nl}}
  }

\begin{document}

%align layout with JAD layout
%\linenumbers %add line number to the text
\doublespacing %add double spacing to the text
\setlength{\voffset}{0in} %set margin on top/bottom of page to 1 inch (weirdly margin = 1 + voffset)
\setlength{\hoffset}{0in} %set margin on left of page to 1 inch (weirdly margin = 1 + hoffset)

\maketitle

\bigskip
\begin{abstract}
\noindent The early detection of Alzheimer’s disease (AD) requires an understanding of the relationships between a wide range of features. Conditional independencies and partial correlations are suitable measures for these relationships, because they can identify the effects of confounding and mediating variables. This article presents a Bayesian approach to Gaussian copula graphical models (GCGMs) in order to estimate these conditional dependencies and partial correlations. This approach has two key advantages. First, it includes binary, discrete, and continuous variables. Second, it quantifies the uncertainty of the estimates. Despite these advantages, Bayesian GCGMs have not been applied to AD research yet. In this study, we design a GCGM to find the conditional dependencies and partial correlations among brain-region specific gray matter volume and glucose uptake, amyloid-beta levels, demographic information, and cognitive test scores. We applied our model to $1\,022$  participants, including healthy and cognitively impaired, across different stages of AD. We found that aging reduces cognition through three indirect pathways: hippocampal volume loss, posterior cingulate cortex (PCC) volume loss, and amyloid-beta accumulation. We found a positive partial correlation between being woman and cognition, but also discovered four indirect pathways that dampen this association in women: lower hippocampal volume, lower PCC volume, more amyloid-beta accumulation, and less education. We found limited relations between brain-region specific glucose uptake and cognition, but discovered that the hippocampus and PCC volumes are related to cognition. These results show that the use of GCGMs offers valuable insights into AD pathogenesis.
\end{abstract}

\vfill

\newpage

\newpage

\spacingset{1.9} % DON'T change the spacing!
%\tableofcontents '

\section*{\textbf{Introduction}}
%add Hoff 2007 introduction about regression models

\label{sec:intro}

Alzheimer’s disease (AD) is clinically characterized by amyloid-beta and tau accumulation. AD is the primary cause of dementia, a condition defined by a decline in cognitive and executive functions \cite{Lancet2024}. Despite extensive research, underlying causes of pathological changes in AD remain unknown \cite{Breijyeh2020}. 

Traditional research in AD typically selects a set of (dependent) variables and assesses the statistical associations with other variables (predictors) \cite{Nemy2020,Teipel2022}. Although this approach can lead to valuable insights, it overlooks the complex network of relations underlying AD pathogenesis. For instance, predictor variables may influence each other (a phenomenon known as collinearity), or a confounding variable may obscure the true relationship between predictors and dependent variables, potentially leading to unreliable results.

Beyond this traditional approach, literature therefore increasingly treats relations between variables from a network perspective. A popular example of this perspective is the field of brain connectivity, which studies the network of relationships between different brain regions \cite{Sporns2013}. Brain connectivity has three subfields: structural, functional, and effective connectivity. Structural connectivity uncovers the physical relations among brain regions \cite{Sporns2005}, for instance using Diffusion Tensor Imaging \cite{Donnell2011}. Functional connectivity involves the covariance, correlation, or other statistical dependency among brain regions\cite{Warnick2018}. Effective connectivity discovers the causal relations among brain regions \cite{Cao2021}; a common statistical technique to study effective connectivity is Dynamic Causal Modeling \cite{Daunizeau2009}.

The present article studies the network of associations between regional gray matter volume and glucose metabolism, but goes beyond the standard brain connectivity approach by also including demographic factors, cognitive scores, and the global amyloid-beta accumulation. In total, we included 19 features that have been shown to play a role in the pathogenesis of AD. Network models between these variables are not new in AD literature and are predominantly causal of nature. A popular tool for causal inference is Structural Equation Modeling (SEM), a family of statistical techniques that, given the causal relations between variables, can estimate the effect size of each relationship \cite{Kline2023}. Mediation analysis is part of the SEM family and studies the contribution of a third variable to the causal dependency between two other variables. It is a common technique in AD research \cite{Caffo2008,Yingxuan2021,Fan2024}. Other applications of SEM in AD include path analysis \cite{Oudenhoven2022}, factor analysis \cite{Neumann2022}, and structural regression \cite{Neumann2022b}. For causal inference in AD research, there are various alternatives to SEM including Structural Causal models \cite{Shen2020}, System Dynamic models, \cite{Uleman2023} and Dynamic Causal Modeling \cite{Daunizeau2009}. 

This article considers undirected graphs. The resulting relations are therefore not causal of nature, as opposed to the directed and causal relations studied by SEMs. Undirected graphs have three main advantages over classical SEMs. First, undirected networks allow cycles. Often a variable A causes B and B causes A. We refer to such relation as a cycle. Cycles are common in AD pathology. SEMs, however, assume the network is acyclic and are therefore not suitable for inference on cyclic relations. Second, causal relations are hard to detect and generally need large sample sizes and longitudinal data. These data are costly to obtain. Inference on undirected networks is possible with smaller sample sizes and does not require longitudinal data. Lastly, undirected networks do not require prior knowledge about the structure of the graph. They can, therefore, be used without limitations on the structure. SEMs do need a causal relation graph as input. SEMs are able to refine the causal structure of this input graph, but these refinements suffer from a poor performance \cite{Shen2020}. SEMs are therefore often restricted to variables and relations of which the causal structure is available a priori. Undirected networks, like the graphical model estimated in this article, can facilitate to form hypotheses about the structure of a causal network. Thus, they have the potential to improve the performance of SEMs and other causal inference methods.

Formally, in an undirected network, an edge between two variables represents some measure of statistical dependency. A common choice for this dependency is the Pearson's correlation or the covariance. However, these metrics can lead to spurious associations by overlooking confounding or mediating factors. Ideally one wants a correlation that is corrected for such factors. Such a correlation exists and is called the partial correlation. Like the Pearson correlation, the partial correlation takes on a value in the range from $-1$ to $1$. Under the assumption that all variables are multivariate normally distributed, a zero partial correlation implies conditional independence. Two variables A and B are conditionally independent when, given all other variables, there is no relation between A and B. Or, put differently, if we keep all other variables fixed, knowing the value of variable A, does not give any information about the value of variable B. One can depict conditional dependencies in an undirected graphical model, where nodes denote random variables and an edge between two nodes/variables is included if and only if the variables are conditionally dependent \cite{lauritzen1996graphical, koller2009probabilistic}. Such graphical models depict confounding and mediating pathways
between large numbers of variables in one figure. They constitute, therefore, a powerful tool to uncovering the complex associations involved in AD, as previously demonstrated \cite{Dyrba.2020}. 

%A popular example that illustrates the power of conditional dependence and partial correlations is the Pearson correlation between ice cream sales and sun burns. Although unrelated, these variables are Pearson correlated, because they are both directly related to high temperatures (the confounder). Accounting for the effect of temperature however, these variables are conditionally independent and have a partial correlation of zero. 

Conditional dependencies, partial correlations, and graphical models are commonly estimated using a frequentist approach. This approach can estimate partial correlations and predict whether any two variables are conditionally dependent or not. In the field of brain connectivity in AD, there are two articles applying frequentist graphical models to uncover conditional dependencies between brain region-specific gray matter volume and glucose uptake\cite{Ortiz2015,Titov2017}. Although frequentist approaches are relatively straightforward, they are not able to provide any uncertainty around the estimated conditional dependencies and partial correlations, a concept referred to as model uncertainty. To overcome this limitation, we employed a Bayesian approach, which offers a key advantage: it not only estimates conditional (in)dependence and partial correlations but also quantifies the uncertainty of these estimates. 

Quantifying this uncertainty is important. Conditional independence, and statistical associations in general, are not black and white; based on the data, relations can be opaque. Consider, for example, the ambiguous role that sex plays in the pathology of AD \cite{Mielke2018}. When answering the question ``Are two variables conditionally dependent?", a simple yes or no might overlook the complexity of the relation. Instead, the answer ``There is conditional dependence with 75\%", gives more color. Uncertainty also allows researchers to state that, based on the data, it is not clear if some variables are conditionally dependent. Quantifying uncertainty becomes even more pivotal if conditional dependence graphs are used as a starting point to develop a causal model, in which mistakes in the starting graph might trickle down to spurious causal results \cite{Shen2020}. 

Despite the advantage that model uncertainty provides, the Bayesian approach is still uncommon in the field of AD research. This is largely due to the dated notion that the Bayesian approach is complex and inefficient. Although this claim is historically true, recent advancements have made Bayesian methods more computationally efficient \cite{mohammadi2024,Vogels2024}, making their application to AD research possible. For example, Bayesian methods have been applied to estimate conditional dependence between brain regions \cite{Dyrba.2018, Dyrba.2020} and to determine the uncertainty of functional connectivity estimates \cite{Hinne2015}. All these applications, however, used so-called Gaussian graphical models (GGMs). In GGMs, all variables are assumed to follow a multivariate normal distribution. In the present study, we employed Gaussian copula graphical models (GCGMs). This methodology enabled the integration of non-normal variables, such as discrete (\textit{e.g.}, amyloid-beta levels), binary (\textit{e.g.}, sex) and continuous variables (\textit{e.g.}, MRI, FDG-PET). 

This study presents a Bayesian approach to GCGMs. Our methodology has three capabilities. i) the estimation of an undirected network depicting the conditional dependencies and partial correlations between variables, ii) the quantification of the uncertainty of these estimates and, iii) the inclusion of all types of variables whether normally distributed or not. Bayesian GCGMs have been successfully applied in other domains \cite{dobra2011copula,Reza2016,Chandra2024}, but remain, to the best of our knowledge, unexplored in the AD domain. 

The application of GCGMs to AD, allowed us to study the conditional dependence structure among 19 relevant features. Specifically, our primary objective was to identify the conditional dependence pathways between demographic variables and cognition (\textit{i.e.}, memory and executive function scores), and between neuro-imaging variables (\textit{i.e.}, gray matter volume, glucose uptake along with amyloid-beta levels) and cognition. Furthermore, we explored the conditional dependency pathways through which demographic variables influence gray matter volume and glucose uptake, and investigated the conditional dependencies between brain region-specific volume and glucose metabolism.

\section*{\textbf{Materials and Methods}}

\subsection*{\textit{Subjects}}
\label{sec:subjects}
We obtained the data for this study from the Alzheimer's Disease Neuroimaging Initiative (ADNI), which provides a public database for AD research including clinical, neuropsychological, neuroimaging, and biomarker data. A complete description of the ADNI and up-to-date information is available at \url{https://adni.loni.usc.edu}. For this study, we selected the baseline examinations of the ADNI-GO and ADNI-2 phases. The final dataset included $1\,022$ participants selected based on the availability of concurrent T1-weighted structural MRI, FDG-PET, amyloid-sensitive AV45-PET data, neuropsychological assessments, and blood-based APOE4 genotyping. 

After initial quality control and data preparation, the final sample included 345 cognitively normal control subjects (CN), 297 patients with amnestic early mild cognitive impairment (EMCI), 205 patients with amnestic late mild cognitive impairment (LMCI), and 175 patients with AD. We included two cognitive composite scores assessing memory (ADNI-MEM) and executive function (ADNI-EF). ADNI-MEM is a weighted average of seven different memory tests \cite{Crane.2012}. Similarly, ADNI-EF combines eleven different executive functions tests in a single score \cite{Gibbons.2012}.
These composite scores provide a higher robustness than individual test scores and a sound psychometric ability to differentiate between different cognitive profiles. Both ADNI-MEM and ADNI-EF range between $-3$ and $3$, a higher score indicating a better performance. Demographics and cognitive test scores of the different diagnostic groups are summarized in Table \ref{tab:sample}.

\begin{table}[h]
\centering
        \begin{tabular}{l r r r r}
            \hline
                                    & {\bf CN}	        & {\bf EMCI}		& {\bf LMCI}        & {\bf AD}	        \\
            \hline
            Sample size	 	& 345 	        & 297 			& 205 			& 175           \\
            Female (\% of total) & $53\%$ & $44\%$ & $43\%$ & $42\%$ \\
            Age (years)				& $74.6 (6.5) $       & $71.6 (7.4)$	    & $74.1 (8.1)$	            & $75.1 (8.0)$	        \\
            Education (years)	     	& $16.5 (2.7)$       & $16.0 (2.7)$	    & $16.2 (2.9)$	            & $15.9 (2.7)$	    \\
            ADNI-MEM	 	& $1.1 (0.60)$	    & $0.58 (0.60)$	    & $0.04 (0.69)$	        & $-0.91 (0.59) $	    \\
            ADNI-EF	 	& $0.83 (0.76)$	    & $0.48 (0.77)$	    & $0.14 (0.87)$	        & $-0.84 (0.90)$\\	    
            \hline
        \end{tabular}
        \caption{\textit{Subject characteristics per diagnosis group. Values denote the mean and standard deviation (in parentheses).}}
    \label{tab:sample}
\end{table}

\subsection*{\textit{Data preparation and feature extraction}}
In this study, we included 19 variables: three demographic variables, two composite cognitive test scores, the number of APOE4 alleles, the amyloid stage, the gray matter volume of six brain regions, and the glucose metabolism of the same six regions. Table S1 in the supplementary material lists all the variables along with their abbreviations and their data type (continuous, discrete, binary or categorical). In this subsection we discuss how we obtained and prepared these variables.

The six brain regions were selected based on our hypotheses and a priori literature findings of early involvement in neurodegenerative processes in AD: hippocampus, caudate, putamen, thalamus, posterior cingulate cortex (PCC), and precuneus \cite{Grothe.2016}. All six regions are defined by the Harvard-Oxford atlas \cite{Desikan.2006}. For each region, we took the average of the left and right brain regions. This avoided the number of variables becoming too large, which would hinder the interpretability of the results. In order to test the robustness of our model, we ran it a second time, this time including the left and right regions separately. 

This study used three biomarkers, all associated with AD:  amyloid-beta accumulation \cite{Lancet2024}, glucose metabolism \cite{Butterfield2019}, and gray matter volume \cite{Bolton2023}. They were measured using AV45-PET, FDG-PET, and T1 weighted MRI scans, respectively. The data resulting from these scans were processed as in earlier article \cite{Grothe.2016, Grothe.2017, Dyrba.2020}. The MRI scans were segmented into gray matter, white matter, and cerebrospinal fluid and spatially normalized to an aging/AD-specific reference template using SPM8 (Wellcome Centre for Human Neuroimaging, University College London) and VBM8 (Structural Brain Mapping Group, University of Jena) toolboxes, and the DARTEL algorithm \cite{Ashburner2007}. FDG- and AV45-PET scans were co-registered to the T1 scan and spatially normalized by applying the deformation fields of the T1 scan. To reduce the effect of partial volume signal arising from the low spatial resolution of PET scans, we applied partial volume correction using a three-compartment model and the MRI-derived tissue segments \cite{MullerGartner.1992}. We scaled the regional gray matter volumes proportionally by the total intracranial volume, the regional FDG-PET values by pons uptake, and the regional AV45-PET values by whole-cerebellum uptake. As amyloid-beta spreading has been shown to follow a specific sequence, we captured the amyloid-beta accumulation in a single five-level amyloid score reflecting the global severity of amyloid-beta deposition \cite{Grothe.2017}. Glucose metabolism and gray matter volume were obtained for each of the six brain regions of interest. 

We included the three demographic variables that are well-known to modulate the risk of AD: age, sex, and education \cite{Lancet2024}. Sex is represented as a binary variable, equal to one for women and zero for men. Education is expressed in the number of years of formal education received. The number of APOE4 alleles is strongly associated with amyloid-beta production and AD \cite{Lancet2024} and, therefore, included as a variable in our analysis. We coded this variable as binary, \textit{i.e.}, zero for patients with no APOE4 allele and one for patients with at least one APOE4 allele. Lastly, we included memory and executive function composite scores named ADNI-MEM and ADNI-EF, respectively. 

\subsection*{\textit{Modeling: Bayesian inference in GCGMs}}
We used a Bayesian framework within Gaussian Copula Graphical Models (GCGMs). This framework combines three essential components: graphical models for estimating conditional dependencies and partial correlations, Bayesian methods for determining the uncertainty of those estimates, and the Gaussian copula for accommodating diverse data types (continuous, discrete, and binary). In this section, we provide a detailed explanation of each component.

%Graphical models estimate G and K
Graphical models \cite{lauritzen1996graphical} represent conditional dependencies between variables in the form of a graph $G$, in which each node represents a variable, and edges connect pairs of variables that are conditionally dependent. So-called Gaussian Graphical Models (GGMs), also known as Markov random fields, assume that all variables $Z_1,..,Z_p$ come from a multivariate normal distribution with mean $0$ and unknown covariance matrix $\mathbf{\Sigma}$. A parameter of interest is the unknown precision matrix $\mathbf{K} = \mathbf{\Sigma}^{-1}$ with entries $k_{ij}$, because a simple transformation of the precision matrix $\mathbf{K}$ gives the partial correlations. Moreover, the sparsity pattern of $\mathbf{K}$ directly encodes the conditional dependence structure. That is, $k_{ij} = 0 \iff Z_i \text { and } Z_j \text{ are conditionally independent}$. GGMs leverage this relationship to recover the graphical model $G$.

%Frequentist methods provide point estimates, but Bayesian methods can provide model uncertainy with the posterior
The aim of GGMs is to use observations of the variables to estimate the precision matrix $\mathbf{K}$ and conditional dependence graph $G$. These observations are denoted by the $n \times p$ matrix $\mathbf{Z}$, which contains $n$ observations of each of the variables $Z_1,..,Z_p$. With the data $\mathbf{Z}$, one can estimate the precision matrix $\mathbf{K}$ and conditional dependence graph $G$. This is commonly done using a frequentist approach, such as the graphical lasso \cite{Friedman2008}. This approach renders a single estimate of $G$ and $\mathbf{K}$, also called a point estimate. In contrast, the Bayesian approach estimates an entire distribution called the posterior. It is given by $P(G,\mathbf{K} | \mathbf{Z})$ and denotes the probability, that, given the data $\mathbf{Z}$, the true conditional dependence graph equals $G$ and the true precision matrix equals $\mathbf{K}$. Due to the posterior, we can go beyond frequentist point estimates and make claims such as: ``variable $A$ and $B$ are conditionally dependent with a probability of $60\%$", or ``the partial correlation between $A$ and $B$ is between $0.1$ and $0.2$ with a probability of $90\%$". In other words, the posterior provides model uncertainty.

%The posterior combines the data and the prior. It can be estimated with MCMC algorithms.
Before obtaining any data, a researcher can already have a belief about $G$ and $\mathbf{K}$. For example, based on literature one might expect the presence of an APOE4 allel to be conditionally dependent with the amyloid stage. In Bayesian statistics, such beliefs are captured in a distribution called the prior \cite{Hoff2009}. In GGMs, this prior is denoted by $P(\mathbf{K},G)$. When no prior information is available, one can choose an uninformative prior that deems every conditional dependence equally likely. In this study, we selected a prior probability of conditional dependence of $20\%$ for all variable pairs. Now, the posterior reflects how the data $\mathbf{Z}$ update this prior belief. This is captured in Bayes' formula given by 
\setlength{\abovedisplayskip}{1pt}
\setlength{\belowdisplayskip}{1pt}
\begin{equation*}
    P(G,\mathbf{K} | \mathbf{Z}) \propto P (\mathbf{Z} | \mathbf{K}) P(\mathbf{K},G),
\end{equation*} 
in which $\propto$ denotes equality up to a constant. Notice in Bayes' formula how the posterior is a combination of the data, given by the likelihood $P(\mathbf{Z} | \mathbf{K})$, and the prior belief, given by $P(\mathbf{K},G)$. Estimating the posterior is commonly done with Markov Chain Monte Carlo (MCMC) algorithms. Such algorithms iteratively obtain samples $(G,\mathbf{K})$ of this posterior. These samples can then be used to make a variety of uncertainty claims about $G$ and $\mathbf{K}$. For example, 
if $30\%$ of the sampled graphs contain an edge between variable $A$ and $B$, then the probability of conditional dependence is $30\%$. 

%Copula graphical models
So far, we have assumed that the data $\mathbf{Z}$ come from a multivariate normal distribution. This assumption, however, limits the applicability of GGMs in practical settings, as real-world data often include non-Gaussian variables. For instance, in our study, the dataset comprises non-Gaussian continuous variables, binary variables (APOE4 and sex), ordinal variables (amyloid stage), and discrete variables (age and education). To overcome this limitation, we employed Gaussian copula graphical models (GCGMs). These models can handle mixed variable types while preserving the theoretical advantages of GGMs. They provide a robust representation of conditional dependency structures and have been successfully applied in neuroscience and brain connectivity studies \cite{Solea2020, Chandra2024}. In GCGMs, the observed variables $Y_1,\dots,Y_p$ are transformed into Gaussian variables $Z_1, \dots, Z_p$. The partial correlations and conditional dependence structure is then calculated for these transformed variables $Z_1, \dots, Z_p$. Continuous variables (brain-region-specific glucose uptake, gray matter volume, ADNI-MEM and ADNI-EF), were transformed once, before the start of the MCMC chain, with a semiparametric transformation \cite{Liu2009}. This transformation is such that the conditional dependencies and partial correlations between the transformed variables reflect those of the observed variables \cite{Liu2009}. We treated age, albeit a discrete variable, as a continuous variable too. Discrete or categorical variables (sex, education, APOE4 presence and amyloid stage) were transformed at every iteration in the MCMC algorithm according to the copula framework\cite{hoff2007extending}. The estimated partial correlations and conditional dependencies resulting from this discrete variable transformation reliably represent the underlying structure of observed data, but do not come with a theoretical guarantee. \cite{hoff2007extending}. In this study we used 120,000 MCMC iterations, discarding the first 20,000 iterations as burn-in. Convergence diagnostics confirmed that the remaining 100,000 iterations were sufficient to provide reliable and stable estimates. 

We refer interested readers to a 
more detailed explanation of Bayesian Gaussian copula graphical models \cite{Reza2016}. For guidance on implementing this approach, we refer readers to the R package BDgraph \cite{mohammadi2019bdgraph}. To ensure reproducibility, the R scripts used to produce our results can be found on the GitHub page \url{https://github.com/lucasvogels33/Modeling-AD-Bayesian-GCGM-from-Demographic-Cognitive-and-Neuroimaging-Data}.

\section*{\textbf{Results}}
\label{sec:results}
%overview
This section presents the results. It contains the estimated conditional dependency networks and the corresponding estimated partial correlations, but also showcases the uncertainty of these estimates. We also briefly discuss how the results change as the disease progresses and what the impact is of considering both the left and right side of each brain region.

\begin{figure}[H]
    \centering
    \begin{tabular}[t]{cc}
        \begin{subfigure}[t]{0.5\textwidth}
            \centering
            \includegraphics[width=0.85\textwidth]{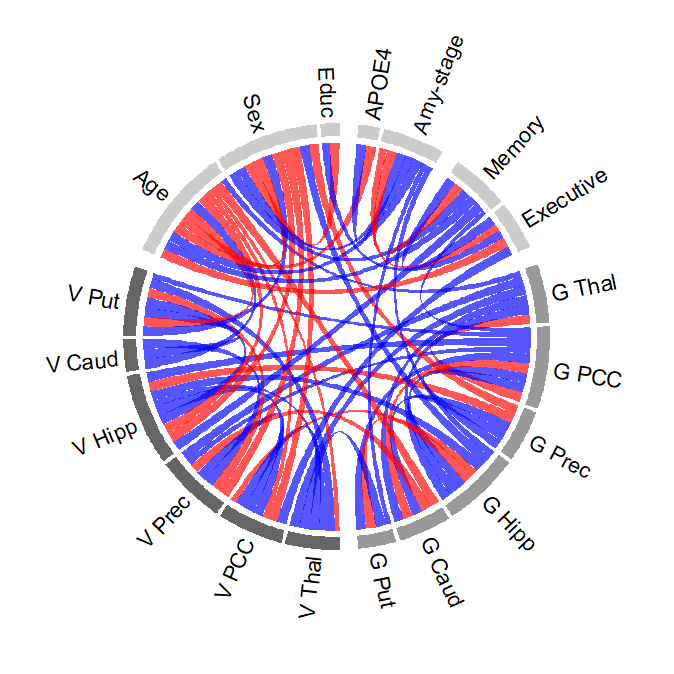}
            \caption{All variables} %{Light Unit}
            \label{fig:network_all}
        \end{subfigure} &
        \begin{subfigure}[t]{0.5\textwidth}
            \centering
            \includegraphics[width=0.85\textwidth]{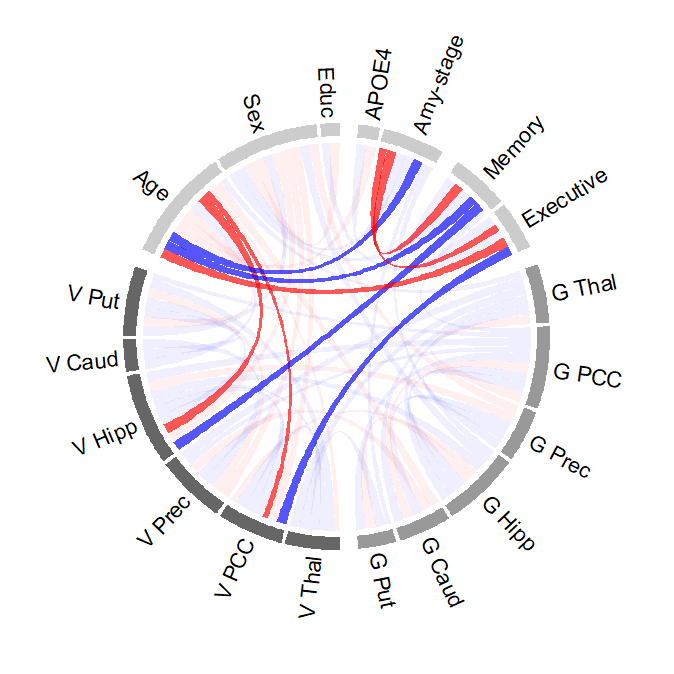}
            \caption{Age and cognition}
            \label{fig:network_age_cog}
        \end{subfigure}
    \\
        \begin{subfigure}[t]{0.5\textwidth}
            \centering
            \includegraphics[width=0.85\textwidth]{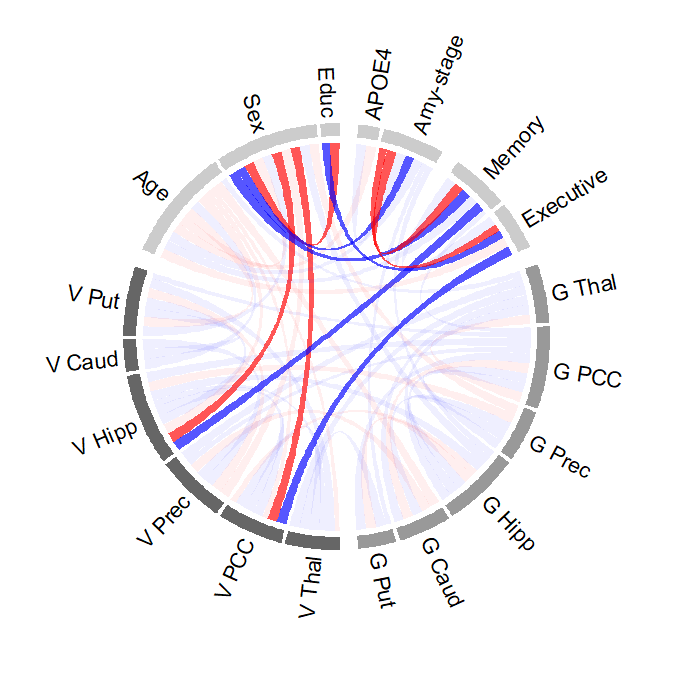}
            \caption{Sex and cognition}
            \label{fig:network_sex_cog}
        \end{subfigure} & 
        \begin{subfigure}[t]{0.5\textwidth}
            \centering
            \includegraphics[width=0.85\textwidth]{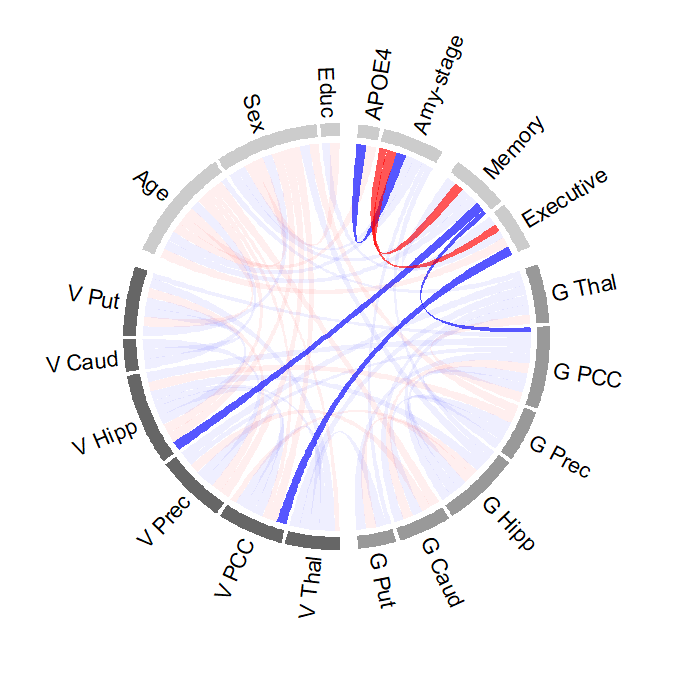}
            \caption{Biomarkers and cognition}
            \label{fig:network_bio_cog}
        \end{subfigure} \\
        \begin{subfigure}[t]{0.5\textwidth}
            \centering
            \includegraphics[width=0.85\textwidth]{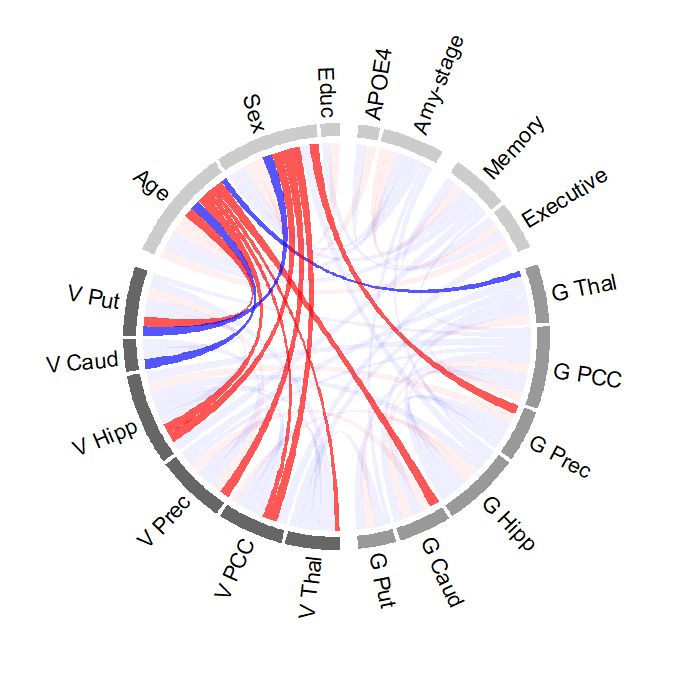}
            \caption{Demograhpics and biomarkers}
            \label{fig:network_dem_bio}
        \end{subfigure} & 
        \begin{subfigure}[t]{0.5\textwidth}
            \centering
            \includegraphics[width=0.85\textwidth]{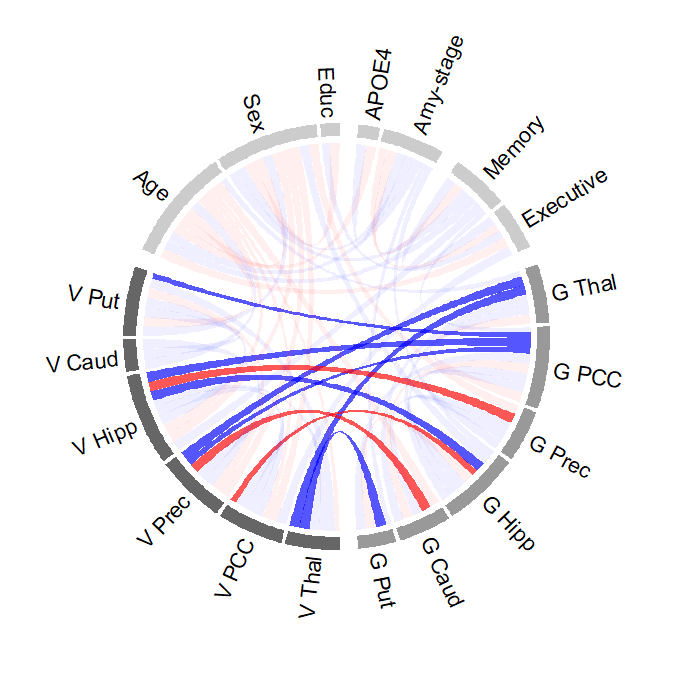}
            \caption{Glucose uptake and volume}
            \label{fig:network_glu_vol}
        \end{subfigure}
    \end{tabular}\\
\caption{\textit{Visualization of conditional dependence among brain-region specific glucose uptake (G), brain-region specific gray matter volume (V), and demographic variables. An edge between variables indicates a conditional dependence with a probability of at least $50\%$. The width of the edges denotes the size of this probability ranging from $50\%$ to $100\%$. A blue (red) edge denotes a positive (negative) partial correlation.}}
\label{fig:network_combined}
\end{figure}

%conditional dependence networks
Figure \ref{fig:network_all} displays all conditional dependencies in a network. In such a network, an edge between a pair of conditionally dependent variables (\textit{e.g.}, age and executive function) is also called a direct pathway. Some pairs of variables are connected via two or more edges. We refer to such connections as indirect conditional dependence pathways. We observe conditional dependency pathways (both direct and indirect), between age and cognition (Figure \ref{fig:network_age_cog}), as well as between sex and cognition (Figure \ref{fig:network_sex_cog}). We observe in Figure \ref{fig:network_bio_cog} that the amyloid stage and brain-region specific gray matter volume are conditionally dependent with cognition, but report limited conditional dependency between brain-region specific glucose uptake and cognition. Figure \ref{fig:network_dem_bio} shows that both old age and being a woman are predominantly negatively partially correlated with brain-region specific volume and glucose metabolism. Lastly, Figure  \ref{fig:network_glu_vol} depicts the ten conditional dependencies between brain-region specific volume and glucose uptake.

\newpage

\setlength{\voffset}{-0.75in}
\setlength{\voffset}{0in}

\begin{figure*}[h]
    \centering
    \begin{subfigure}[b]{0.5\textwidth}
        \centering
        \includegraphics[height=3.2in]{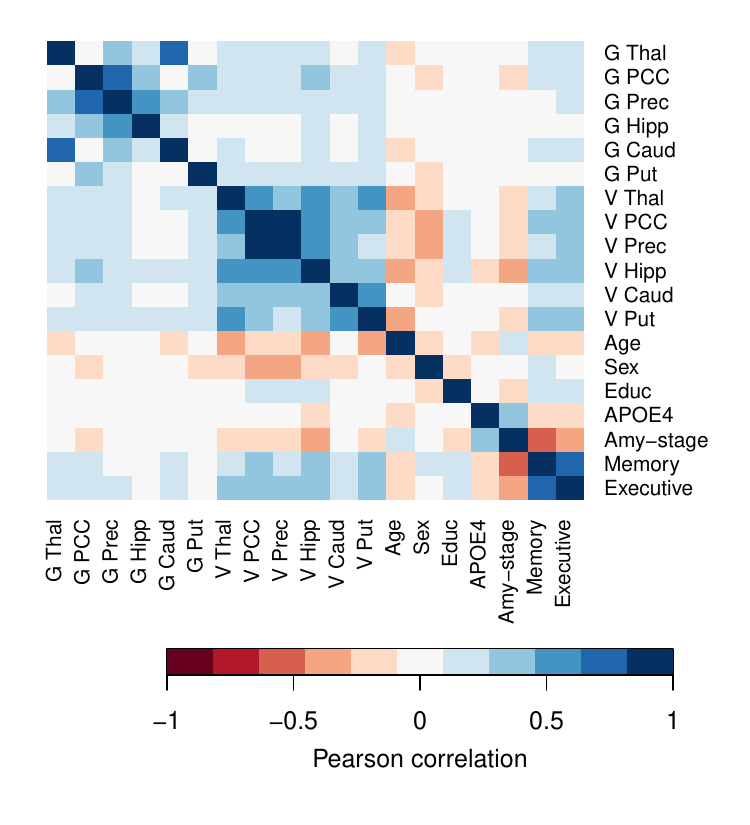}
        \caption{Pearson correlation}
        \label{fig:pearson}
    \end{subfigure}%
    ~ 
    \begin{subfigure}[b]{0.5\textwidth}
        \centering
        \includegraphics[height=3.2in]{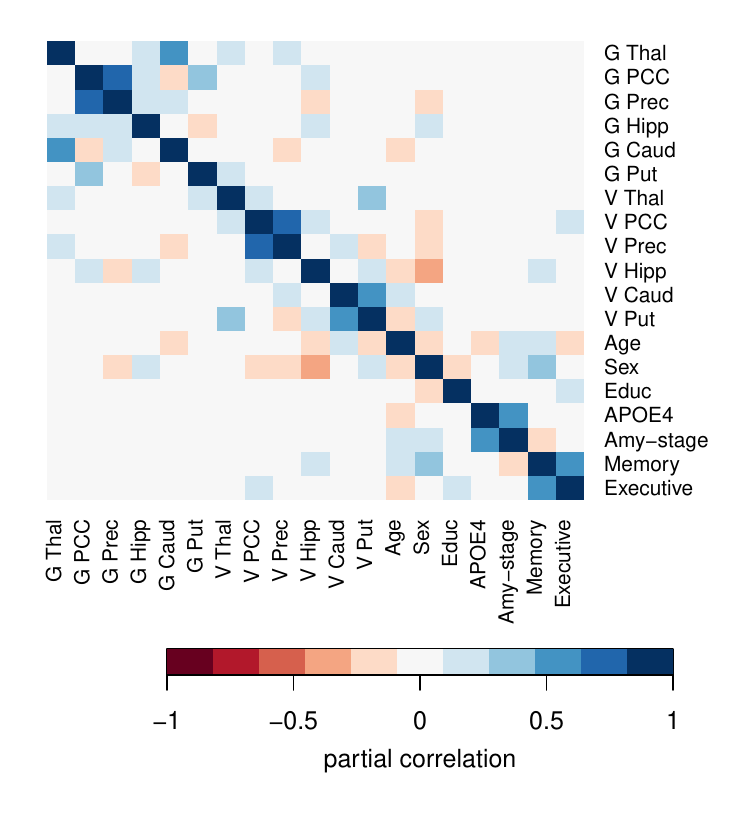}
        \caption{Partial correlation}
        \label{fig:partial}
    \end{subfigure}
    \caption{\textit{Pearson correlations (left) and partial correlations (right) among brain-region specific glucose uptake (G), brain-region specific gray matter volume (V), and demographic variables. Pearson correlations are set to zero when p-value $> 0.05$. Partial correlations are set to zero when the corresponding edge inclusion probability is smaller than $50\%$.}}
    \label{fig:corr_complete}
\end{figure*}

%correlation heatmaps
The conditional dependence networks in Figure \ref{fig:network_combined} reveal what variables were likely to be conditionally dependent. They do not reveal, however, the strength of this dependence. These are given by the partial correlations and are shown, alongside the Pearson correlations in Figure \ref{fig:corr_complete}. Pearson correlations were set to zero when the $p$-value exceeds $0.05$, while partial correlations were set to zero when the corresponding edge inclusion probability is below $50\%$. The Pearson correlation heatmap (Figure \ref{fig:pearson}) is denser compared to the partial correlation heatmap (Figure \ref{fig:partial}). The average absolute Pearson correlation was $0.17$, while the average absolute partial correlation was only $0.07$. Moreover, just $27\%$ of Pearson correlations were set to zero versus $65\%$ of partial correlations. Strong Pearson correlations (with an absolute value greater than $0.25$) were fairly common ($26\%$), while strong partial correlations were less frequent ($7\%$). 

\begin{figure}[H]
     \centering
     \includegraphics[width=0.75\textwidth]{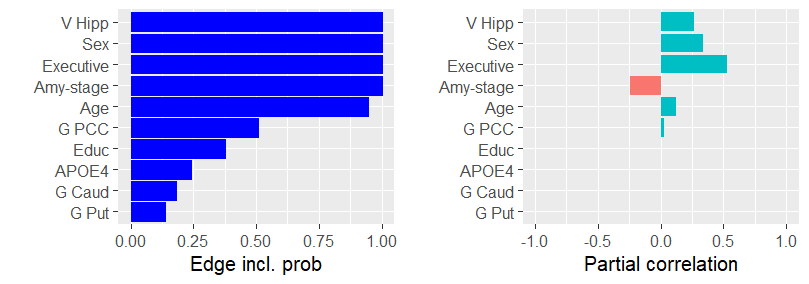}
    \caption{\textit{The probability of being conditionally dependent with memory (left) and the partial correlation with memory (right). }}
    \label{fig:memory_info}
\end{figure}

\begin{figure}[H]
     \centering
     \includegraphics[width=0.75\textwidth]{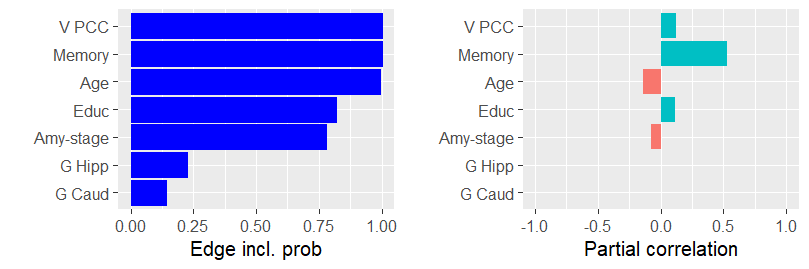}
    \caption{\textit{The probability of being conditionally dependent with executive function (left) and the partial correlation with executive function(right).}}
    \label{fig:ef_info}
\end{figure}

%memory and executive function
We applied GCGMs to AD primarily to discover the conditional dependencies related to cognition (\textit{i.e.}, memory and executive function). Figure \ref{fig:memory_info} presents the ten variables that were most likely conditionally dependent with memory. For each variable, the figure displays the probability of conditional dependence (left) and the mean partial correlation (right). Memory was partially correlated with hippocampus volume ($0.25$), being a woman ($0.3$), amyloid stage ($-0.25$), and age ($0.1$). Figure \ref{fig:ef_info} presents the same information for executive function. Executive function was partially correlated with PCC volume ($0.1$) and age ($-0.1$). For both education and amyloid stage, we reported a $75\%$ probability of conditional dependence with executive function.

\begin{figure}[H]
    \centering
    \begin{tabular}[t]{cc}
        \begin{subfigure}[t]{0.5\textwidth}
            \centering
            \includegraphics[width=0.85\textwidth]{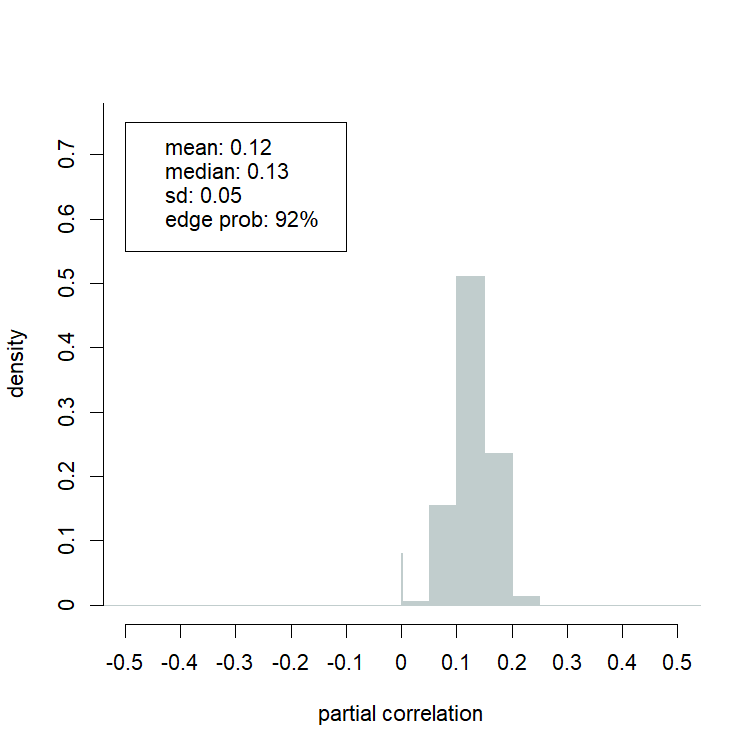}
            \caption{Age and memory}
            \label{fig:density_age_memory}
        \end{subfigure} & 
        \begin{subfigure}[t]{0.5\textwidth}
            \centering
            \includegraphics[width=0.85\textwidth]{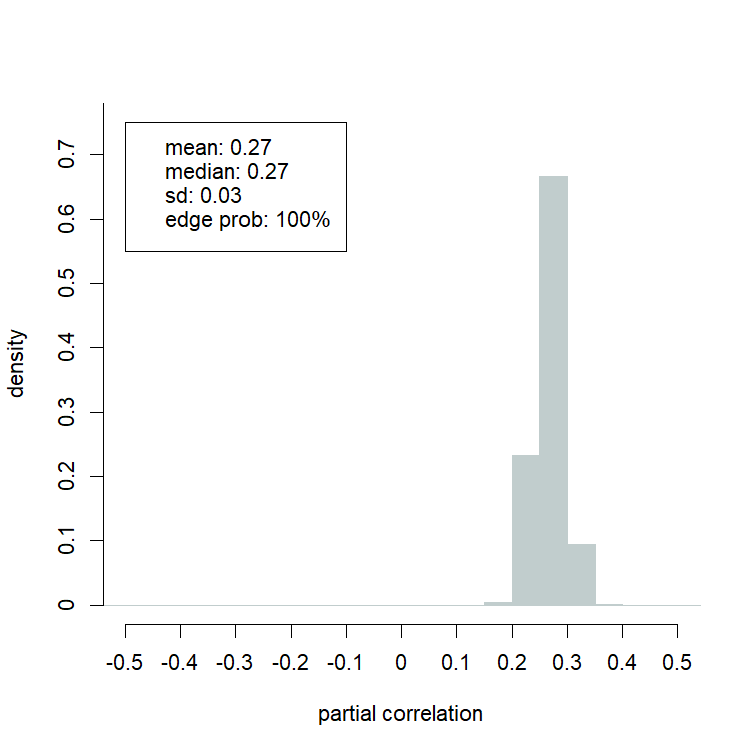}
            \caption{Hippocampus volume and memory}
            \label{fig:density_VHipp_memory}
        \end{subfigure} \\
        \begin{subfigure}[t]{0.5\textwidth}
            \centering
            \includegraphics[width=0.85\textwidth]
            {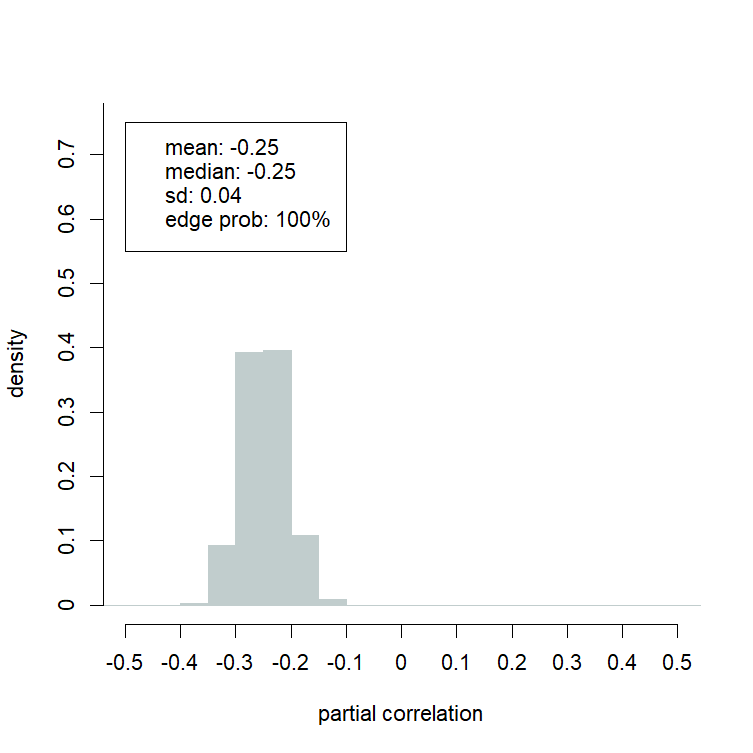}
            \caption{Amyloid stage and memory}
            \label{fig:density_amstage_memory}
        \end{subfigure} & 
        \begin{subfigure}[t]{0.5\textwidth}
            \centering
            \includegraphics[width=0.85\textwidth]{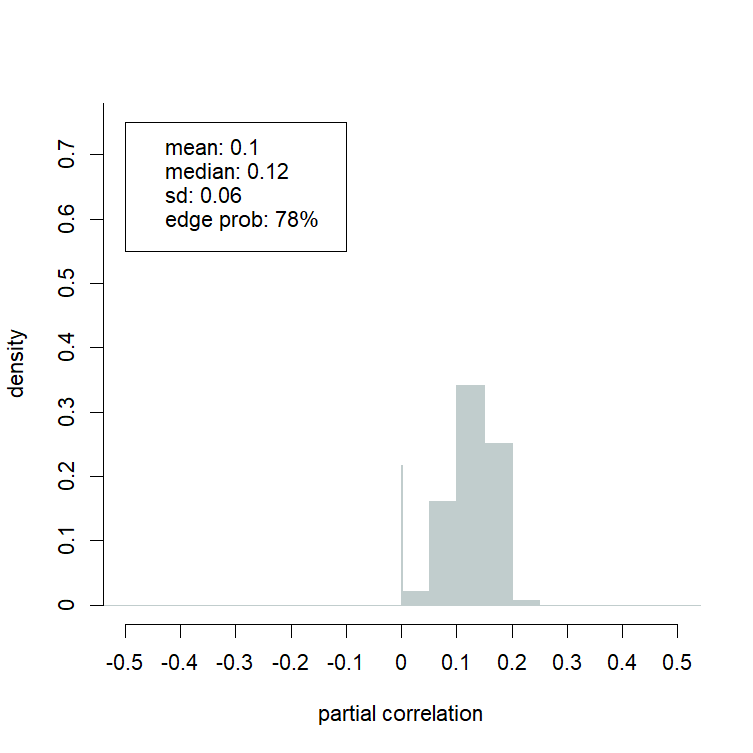}
            \caption{Executive function and education}
            \label{fig:density_educ_executive}
        \end{subfigure}
    \end{tabular}\\
\caption{\textit{Estimated probability density plots of the partial correlations of four selected pairs of variables.}}
\label{fig:densities}
\end{figure}

%uncertainty
The main advantage of the Bayesian approach is its ability to compute the uncertainty of its estimates. We showcase this in Figure \ref{fig:densities} which presents the probability density plots of four selected partial correlations. The spike at zero in some of these plots denotes the probability that the partial correlation is zero, \textit{i.e.}, the probability that the variables were conditionally independent. The partial correlations depicted in the plots have low standard deviations ($\leq 0.06$). In fact, across all pairs of variables, the average standard deviation was just $0.02$ with a maximum of $0.06$. The estimated means of the partial correlations, as reported in Figure \ref{fig:partial}, were therefore likely to be close to the true partial correlations.  

\setlength{\voffset}{-0.75in}
\setlength{\voffset}{0in}

%disease stage specific
So far, we have considered the results of $1\,022$  patients combined over all four disease stages: healthy, EMCI, LMCI, and AD. Next, we examined how pathways vary across different disease stages by analyzing each stage separately. The resulting networks are shown in supplementary material S2. We observe that the disease-stage specific networks are sparser than the combined model, with approximately $80\%$ of variable pairs being conditionally independent in each of the four stage specific models, compared to $65\%$ in the combined model. The average partial correlation value also decreased from $0.07$ in the combined model to around $0.05$ in each of the disease-stage specific models. This increased sparsity can be attributed to reduced within-group variance. This reduced variance made it harder for the model to deduce conditional dependencies. The reduced variance was observed in all variables, but was particularly present in the memory and executive function variables. The composite memory score, for example, ranged between $-3$ and $3$ in the combined model. Among AD patients, however, it ranged between $-3$ and $0.5$. This is why, for this disease stage, memory and executive function had almost no conditional dependencies left, see Figure S1(e) in the supplementary material. In later disease stages, reduced variance also caused age and sex to lose conditional dependencies with APOE4 and the amyloid score. The hippocampus volume was an exception to this trend and had more conditional dependencies as the disease progresses. Of the pathways that we found for the combined model, some continued to exist in all four disease stages. Examples are the conditional dependencies between age and brain volume, as well as between sex and brain volume. Other pathways persised only in one disease stage, predominantly in the EMCI stage. These included the pathway linking age to memory via the amyloid stage, or the pathway linking sex and memory through the hippocampus volume. 

%left and right regions
We also conducted a separate analysis in which we included the left and right brain regions separately. The resulting network with $31$ variables is shown in the supplementary material Figures S2 and S3 and was sparser compared to the original, merged model. Specifically, $77\%$ of conditional dependencies had a lower probability than 0.5, compared to $65\%$ in the combined model. Additionally, the average partial correlation decreased to $0.047$ from $0.07$ in the original model. Our previous conclusions remain largely unchanged. Notably, sex continued to influence memory through education, amyloid-beta, hippocampal volume (V Hipp), and posterior cingulate cortex volume (V PCC), while age still impacted memory through amyloid-beta. However, some associations no longer held: age was no longer conditionally dependent with hippocampal volume, and amyloid-beta no longer associated with executive function. Separating the left and right brain region analysis, also provided extra insights: the volumes of the left and right brain regions were highly partially correlated, as is their glucose metabolism. Furthermore, the left hippocampal volume was specifically linked to memory, while the left PCC showed a gender-related association. Finally, the association between volume and metabolism was notably stronger within the same brain regions.

\section*{\textbf{Discussion}}
This section interprets the results of the previous section and compares them with the existing results in the literature. We also discuss the limitations of our work and end with a conclusion. 

\subsection*{\textit{Interpretation of results}}
The interpretation of all conditional (in)dependencies and partial correlations of section 3 goes beyond the scope of this paper. Instead, we interpret here those associations that we deem relevant. 

We first elucidate the conditional dependencies (direct and indirect) between the demographic variables (age, sex, and education) and cognition (memory and executive function), starting with age. It is well-established in the literature that age is the biggest risk factor for AD \cite{Lancet2024}. In line with this, we found negative Pearson correlations between age and cognition (Figure \ref{fig:pearson}) and a negative partial correlation between age and executive function (Figure \ref{fig:ef_info}). However, we found a surprising positive partial correlation between age and memory. Specifically, we estimated that age and memory are conditionally dependent with $92\%$ and that their partial correlation is between $0.05$ and $0.2$ with a $90\%$ probability (Figure \ref{fig:density_age_memory}). At first glance, this result suggested a possible limitation in the model. However, we found a similar result using a linear regression using the same $19$ variables of our model with memory as a dependent variable: the resulting regression coefficient of age was positive ($0.011$) and significant ($p$-value $<0.001$). The positive partial correlation and regression coefficient suggested that there was a confounding variable not in our model through which age positively associated with memory. A potential candidate for this confounder is lifestyle. As noted in literature, $45\%$ of AD cases can be attributed to 14 lifestyle related risk factors \cite{Lancet2024}, of which only education was included in our model. The older patients in our dataset might be the ones with a better lifestyle, and therefore a better memory.

The positive partial correlation between age and memory also suggested that the main indirect pathways through which age contributes to memory decline were captured in our model. Figure \ref{fig:network_age_cog} shows two such pathways. Aging increased amyloid-beta deposition and reduced the gray matter volume of the hippocampus and PCC, impairing memory and executive function. Both associations were confirmed in the literature \cite{Lancet2024,Ma2022,Pini2016}. The GCGM model allowed us to estimate the uncertainty of these estimates. For example, with more than $95\%$ certainty we know that the partial correlation between the amyloid stage and memory was between $-0.15$ and $-0.35$ (Figure \ref{fig:density_amstage_memory}). Similarly, we observed that with almost $100\%$ certainty the partial correlation between memory and the gray matter volume of the hippocampus ranged between $0.2$ and $0.35$ (Figure \ref{fig:density_VHipp_memory}).

Next, we look at the relation between sex and cognition. We found a positive partial correlation between being a woman and memory (Figure \ref{fig:partial}). The Pearson correlation between sex and memory was weaker (Figure \ref{fig:pearson}). This suggested that there may exist indirect conditional dependency pathways between sex and cognition, each of which dampening the positive effect of the observed direct partial correlation. Figure \ref{fig:network_sex_cog} depicts four such pathways. First, women in our dataset have spent, on average, one year less in education. This effect was also observed in literature \cite{Rocca2017,Castro2023,Lancet2024}. The density plot in Figure \ref{fig:density_educ_executive} suggested that education and executive function were conditionally dependent with a probability of $78\%$ with a partial correlation ranging between $0$ and $0.2$. Second, we observed a positive partial correlation between being a woman and amyloid accumulation. An association also found by others\cite{Nemes2023}. This association may provide evidence that amyloid-beta deposition is associated with our brain’s immune system \cite{Eimer2018}, which was reported to be stronger in women \cite{Klein2016}. Third, we found that women have smaller hippocampal volumes. A conclusion also found in a large study with $18\,600$ individuals\cite{Dima2022}. Lastly, we found a negative partial correlation between sex and PCC gray matter volume. We found a $40\%$ chance that memory was conditionally dependent on education (Figure \ref{fig:memory_info}) and a $78\%$ chance that executive function was conditionally dependent on education (Figure \ref{fig:ef_info}). This is in line with the prevailing notion that education improves cognition and reduces the risk of developing AD \cite{Lancet2024}. 

Next, we identified the conditional dependency pathways between neuro-imaging variables (gray matter volume, glucose uptake and amyloid-beta accumulation) and cognition. Figure \ref{fig:network_bio_cog} depicts all such dependencies. We reported a conditional dependency between the hippocampal volume and memory with a probability of one, see Figure \ref{fig:memory_info}. This supported the established notion that the hippocampus is associated with memory \cite{Ali2016}. Figure \ref{fig:ef_info} shows an almost certain conditional dependency between the PCC volume and executive function, confirming a theory that PCC is involved in attention \cite{Hahn2006}. Our findings suggested that the high Pearson correlations between brain-region specific volume and cognition (Figure \ref{fig:pearson}) were solely due to each region's conditional dependence with the hippocampus and PCC volume. Corrected for other variables, we found no relation between glucose metabolism and cognition. The link between brain-region specific glucose uptake and cognition has not been widely studied, although there is evidence for an association between PCC glucose uptake and cognition \cite{Landau2011}. We found a conditional dependence between these two variables with a $50\%$ probability, but the corresponding partial correlation was close to zero (Figure \ref{fig:memory_info}). We observed that the presence of at least one allele of the APOE4 gene was partially correlated with amyloid-beta accumulation, which, in turn, was conditionally dependent with both executive function and memory. This came as no surprise, as this supported the widely accepted hypothesis that the presence of APOE4 alleles is linked to amyloid-beta accumulation and cognitive decline \cite{Ma2022}.

Third, we investigated the conditional dependencies between demographic variables (sex and age) and brain-region specific glucose uptake and volume. Figure \ref{fig:network_dem_bio} depicts such dependencies. We found that age and gender showed more conditional dependencies with brain-region specific volume than with brain-region specific glucose uptake. Age was negatively partially correlated with the volume of the putamen, hippocampus, PCC and thalamus. Age induced atrophy in these regions was also described in literature\cite{Raji2022}. We reported negative partial correlations between being female and the volume of three brain regions: the hippocampus, precuneus and PCC. We also found a positive partial correlation between being a woman and the volume of the putamen, an association also reported by others\cite{Nunnemann2009}. Concerning glucose uptake, age was only negatively partially correlated with the caudate. This aligns with earlier findings that the precuneus cortex, hippocampus, thalamus, and putamen are among the regions whose metabolism is least affected by aging\cite{Berti2014}.

Lastly, we looked at the conditional dependencies between brain-region specific volume and glucose metabolism. They are depicted in Figure \ref{fig:network_glu_vol}. We found ten such conditional dependencies. Figure S4 in the supplementary material shows that these conditional dependencies increased in number as the disease progresses. Among EMCI patients, we found almost no conditional dependencies between brain volume and glucose uptake, whereas in AD patients several such conditional dependencies can be observed. Among AD patients, the volume and glucose uptake of the hippocampus were positively partially correlated. The same was true for the thalamus and the putamen. This was in line with the hypothesis that in AD reduced glucose metabolism precedes neuronal loss and brain volume reduction \cite{Butterfield2019}.

\subsection*{\textit{Limitations}}
Our Bayesian approach to Gaussian Copula graphical models (GCGMs) has three main limitations. First, it becomes slow for practical applications when the number of variables $p$ and the number of observations $n$ increases. The GCGM discussed in this paper had $p=19$ variables and $n=1\,022$ observations and ran within $10$ minutes. A larger model, however, with more than $100$ variables and/or more than $5000$ observations would be infeasible for a Bayesian GCGM. Bayesian uncertainty evaluation relies on an MCMC algorithm that iteratively samples new graphs and precision matrices, both of a dimension $p \times p$. Increasing $p$ therefore exponentially increases the running time and memory requirements of the model. Moreover, at every MCMC iteration, the algorithm needs to resample every non-continuous variable $n$ times. This resampling allows for the inclusion of binary, discrete and ordinal variables, but leads to a lack of scalability in the number of observations. 

The second limitation of the GCGM is its linearity assumption. More specifically, partial correlations represent the strength of a linear relationship with the effect of other variables removed. Partial correlations can be over-, or underestimated, when the underlying relationship is not linear. 

Lastly, GCGMs are undirected. This comes with advantages. GCGMs can deal with cyclic networks, require relatively few observations and do not need prior knowledge of the structure. However, the undirected nature of GCGMs make them not suited for identifying causal effects, such as the impact of a treatment on amyloid-beta accumulation, or the effect of a lifestyle change on cognition. The undirected networks provided by GCGMs can, however, be used as a starting point for causal analysis.

The particular application of GCGMs described in this article also comes with three caveats, each opening an avenue for further research. First, apart from education, we did not include any lifestyle variables in our model. These variables account for an estimated $45\%$ of AD cases\cite{Lancet2024} and are a probable cause for the surprising positive partial correlation between old age and memory. Second, Bayesian models allow for the inclusion of prior knowledge. For simplicity, we choose in this article for an uninformative prior in which every pair of variables is conditionally dependent with $20\%$. Future work could use AD literature to construct an informed prior, certainly for variable pairs whose association is known, such as APOE4 and the amyloid stage. The use of an informed prior could lead to a more realistic conditional dependence structure. Lastly, this work did not consider temporal data, and therefore, did not cover the progression of AD over time. Future research could add the disease stage over time as a set of variables. This would allow the model to discover the most important factors contributing to the conversion of CN to EMCI to LMCI, and ultimately, to AD.

\subsection*{\textit{Conclusions}}
In this article we present a Bayesian approach to Gaussian copula graphical models (GCGMs). This approach has three capabilities: i) the estimation of an undirected network depicting the conditional dependencies and partial correlations among variables, ii) the quantification of the uncertainty of these estimates and, iii) the inclusion of all types of variables whether normally distributed or not. Bayesian GCGMs have been successfully applied in other domains \cite{dobra2011copula,Reza2016,Chandra2024}, but remain, to the best of our knowledge, unexplored in the AD domain. 

Our GCGM uncovered the partial correlations and conditional dependencies among demographic data, the global amyloid stage, cognitive test scores, brain-region specific gray matter volume, and brain-region specific glucose uptake. Our Bayesian approach enabled us to estimate the uncertainties of these estimates too. 

As expected, we observed that the partial correlations and their corresponding conditional dependency networks were sparser than the commonly used Pearson correlation or covariance measures. Our study confirmed existing knowledge, but also opened up new hypotheses. We found three indirect pathways through which old age reduces cognition: hippocampal volume loss, PCC volume loss, and amyloid-beta accumulation. Moreover, we found that women performed better on cognitive tests, but also discovered four indirect pathways that dampen this effect: lower hippocampal volume, lower PCC volume, more amyloid-beta accumulation, and less education. We also found that the hippocampus and PCC volumes are conditionally dependent on cognition, but found limited conditional dependence between brain-region specific glucose uptake and cognition. We found that age and sex were more conditionally dependent with brain-region specific volume than with brain-region specific glucose uptake. Lastly, we discovered that the conditional dependence between brain-region specific volume and glucose metabolism increased as the disease progresses.

\section*{\textbf{Acknowledgements}}
Data collection and sharing for this project was funded by the Alzheimer's Disease Neuroimaging Initiative (ADNI) (National Institutes of Health Grant U01 AG024904) and DOD ADNI (Department of Defense award number W81XWH-12-2-0012), and the Alzheimer's Disease Metabolomics Consortium (National Institute on Aging Grants R01AG046171, RF1AG051550 and 3U01AG024904-09S4). The ADNI is funded by the National Institute on Aging, the National Institute of Biomedical Imaging and Bioengineering, and through generous contributions from the following: AbbVie, Alzheimer's Association; Alzheimer’s Drug Discovery Foundation; Araclon Biotech; BioClinica, Inc.; Biogen; Bristol-Myers Squibb Company; CereSpir, Inc.; Cogstate; Eisai Inc.; Elan Pharmaceuticals, Inc.; Eli Lilly and Company; EuroImmun; F. Hoffmann-La Roche Ltd and its affiliated company Genentech, Inc.; Fujirebio; GE Healthcare; IXICO Ltd.; Janssen Alzheimer Immunotherapy Research \& Development, LLC.; Johnson \& Johnson Pharmaceutical Research \& Development LLC.; Lumosity; Lundbeck; Merck \& Co., Inc.; Meso Scale Diagnostics, LLC.; NeuroRx Research; Neurotrack Technologies; Novartis Pharmaceuticals Corporation; Pfizer Inc.; Piramal Imaging; Servier; Takeda Pharmaceutical Company; and Transition Therapeutics. The Canadian Institutes of Health Research is providing funds to support ADNI clinical sites in Canada. Private sector contributions are facilitated by the Foundation for the National Institutes of Health (\url{https://www.fnih.org}).

Lastly, we acknowledge Elias Dubbeldam, PhD student at the Amsterdam Business School, for his feedback on the article. In particular his suggestions on the visualization of the conditional dependencies improved the article.

\section*{\textbf{Funding}}
The authors received no financial support for the research, authorship, and/or publication of this article.

\section*{\textbf{Conflict of interest}}
For this research we collaborated with Stefan Teipel. He is an editor of the supplemental issue on the early detection of Alzheimer's to which this manuscript is submitted. This forms a potential conflict of interest. The authors declared no other potential conflicts of interest with respect to the research, authorship, and/or publication of this article.

\section*{\textbf{Data availability statement}}
The data supporting the findings of this study are available upon request from the corresponding author. The data are not publicly available due to privacy and ethical restrictions. The R script used to produce our results can be found on the GitHub page \url{https://github.com/lucasvogels33/Modeling-AD-Bayesian-GCGM-from-Demographic-Cognitive-and-Neuroimaging-Data}.

\bibliographystyle{unsrt.bst}
\bibliography{sampleBIB}
\end{document}